\def\P{ \tilde{\rm P}^\emptyset_i}
\def\Q{ \tilde{\rm P}^\emptyset_j}
\def\T{ \tilde{\rm T}^{\emptyset +}_i}
\def\TT{ \tilde{\rm T}^{\emptyset +}_j}
\def\etal{ et al.}
\def\ber{\begin{eqnarray}}
\def\eer{\end{eqnarray}}
\def\etal{et al}
\def\be{\begin{equation}}
\def\ee{\end{equation}}

\documentclass[aps,prb,showpacs,superscriptaddress,a4paper,twocolumn]{revtex4-1}
\usepackage{amsmath}
\usepackage{float}
\usepackage{subfigure}
\usepackage{color}
\usepackage{color, colortbl}
\definecolor{Gray}{gray}{0.75}
\definecolor{LightCyan}{rgb}{0.50,1,1}
\usepackage[section]{placeins}
\usepackage{amsfonts}
\usepackage{amssymb}
\usepackage{graphicx}
\begin {document}
\author{Banasree Sadhukhan}\email{banasree.rs@presiuniv.ac.in}
\author{Arabinda Nayak}\email{arabinda.physics@presiuniv.ac.in}
\affiliation{Department of Physics, Presidency University, 86/1College Street, Kolkata 700073, India}
\author{Abhijit Mookerjee}\email{abhijit@bose.res.in}
\affiliation{Professor Emeritus, S.N. Bose National Center for Basic Sciences, JD III, Salt Lake, Kolkata 700098, India}

\title{\bf Effect of disorder on the optical response of NiPt and Ni$_3$Pt alloys.}
\date{\today}
 
\begin{abstract}
In this communication we present a detailed study of the effect of chemical disorder on the optical response of Ni$_{1-x}$Pt$_x$ (0.1$\leq$ x $\leq$0.75) and Ni$_{3(1-x)/3}$Pt$_x$ (0.1$\leq$ x $\leq$0.3). We shall propose a formalism which will combine a Kubo-Greenwood approach with a DFT based tight-binding linear muffin-tin orbitals (TB-LMTO) basis and augmented space recursion (ASR) technique to explicitly incorporate the effect of disorder. We show that chemical disorder has a large impact on optical response of Ni-Pt systems. In ordered Ni-Pt alloys, the optical conductivity peaks are sharp. But as we switch on chemical disorder, the UV peak becomes broadened and its position as a function of composition and disorder carries the signature of  a  phase transition from NiPt to Ni$_3$Pt with decreasing Pt concentration. Quantitatively this agrees well with Massalski's Ni-Pt phase diagram \cite{massal}. Both ordered NiPt and Ni$_3$Pt have an optical conductivity transition at 4.12 eV. But disordered NiPt has an optical conductivity transition at 3.93 eV. If we decrease the Pt content, it results a chemical phase transition from NiPt to Ni$_3$Pt and shifts the peak position by 1.67 eV to the ultraviolet range at 5.6 eV. There is a significant broadening of UV peak with increasing Pt content due to enhancement of 3d(Ni)-5d(Pt) bonding. Chemical disorder enhances the optical response of NiPt alloys nearly one order of magnitude. Our study also shows the fragile magnetic effect on optical response of disordered Ni$_{1-x}$Pt$_x$ (0.4$<$ x $<$0.6) binary alloys. Our theoretical predictions agree more than reasonably well with both earlier experimental as well as theoretical investigations.
\end{abstract}

\date{\today}

\maketitle


\section{Introduction : Motivation}
         Transparent conductive electrodes are  indispensable
components of many opto-electronic devices including solar
cells, organic light emitting diodes, photodetectors, touchscreens, photovoltaics, integrated modulators, liquid-crystal displays, electrochromic devices and photocathodes of dye
synthesized solar cells \cite{solar}. There has been extensive research interest in producing high quality transparent conductive oxides (TCOs) \cite{Chopra}-\cite{jarz} in which indium tin oxide (ITO) \cite{Laux}-\cite{Amaral} has been the most widely used. Alternatives include other transparent conducting oxides, such as aluminium doped zinc oxide (ZnO:Al) \cite{zno}-\cite{zno1}, fluorine doped tin oxide, conducting polymers, carbon nanotubes, graphene, and semiconductor nanowires. ITO has the drawback of being expensive and is not compatibile with organic materials \cite{Jiang}. Recent research has concentrated in producing alternatives. The problems have been overcome in ultra thin metallic films (UTMF) \cite{giur},\cite{ghosh1} based on transition metals  \citep{granqvist},\cite{ghosh}. A potential drawback of UTMFs is  degradation due to oxidation. In fact, contact with other reactive chemical elements can also take place during fabrication (e.g. wet etching). UTMFs have always run the risk of changing their electrical and optical properties, in particular, layers of chromium, nickel, titanium and aluminum. Of course, single-component film based on transition metals, like Cr and Ni, can overcome this drawback. Ni films \cite{marti} deposited by single-step sputtering can be effective transparent electrodes over the entire wavelength ranging from the ultra violet (175 nm) to the  mid infra-red  (25 $\mu$m). In fact a detailed comparison with ITO indicates a similar performance in the visible range, while significant improvement is found in the UV and IR regions. The measured wide optical transmission and excellent electrical properties, combined with the proven stability after an ad hoc passivation treatment, make Ni based UTMFs serious competitors to TCOs, in particular in the UV and IR ranges. Recently, Pd (4d) and Pt (5d) \cite{wang} rich alloy of 3d transition metals showes interesting optical and magnetic properties which leads to a variety of applications \cite{cadev}-\cite{parcab}. Thin  film  of those alloys can replace both TCO and UTMF. Ni based thin alloy-films are far better than single component Ni based UTMF. The optical as well as magnetic properties of alloy films can be tailored according to our requirements.

In the start of any scientific project we have to choose a particular problem. This choice is crucial if
we wish to make any  significant contribution.This extended introduction was necessary to justify our decision to study the effect of disorder on alloy systems. Our particular choice of application of our methodology on Pd and Pt based Ni alloys required this justification. However, the very importance of these materials has lead to a large body of published  theoretical work to understand their behaviour. We still have to justify why we wish to present one more approach. If it is only for the sake of presenting a slightly different methodology, then the new understanding will be incremental and not worth the trouble. At the outset we wish to state that our aim is different.  We want to present a methodology which will be applicable to a large class of disordered systems : chemical and structural disorder, systems which exihibit short ranged order like clustering and segregation, as well as long-ranged disorder like random chains of Stone-Wales defects in graphinic solids \cite{SW}. It should be able to explain the behaviour of metallic and insulating solids as well as solids with magnetic ordering. Moreover, our idea is not only gaining qualitative understanding but also obtaining quantitative predictions which can be `proved' by experiment. If we wish to tailor materials to our specific needs, we must theoretically understand the material and its response properties accurately. With this in hand, the experimentalist can  begin  with confidence and much greater chance of success. The aim of this work is to study, in detail, the optical response in Ni-Pt alloys.

                In two earlier works by Wang \etal \cite{wang} and our group \cite{durga}, it was shown that the stability of NiPt alloys are higher than those of NiPd due to predominance of relativistic effects \cite{wang}-\cite{durga}. Wang and Zunger \cite{wang} have shown that with the inclusion of relativisic corrections in formation energies,  ordered NiPt is always more stable than the corresponding NiPd with the same alloy composition.  In our previous investigation \cite{durga} we had come to the same conclusion. Relativistic corrections result in an upshift of the 5d band of one component, bringing it closer to the 3d band of the other. This significantly enhances the 3d-5d hybridization. Relativistic orbital contraction reduces the lattice constant of 5d-Pt  and lowers its size mismatch with 3d-Ni. This results in the reduction of the strain energy associated with the size mismatch between the components. Both the enhanced d-d hybridization and the reduction of packing strain are larger in 3d-5d NiPt than in 3d-4d NiPd. At low temparature NiPt alloys have negative formation energies and form stable, ordered 3d-5d alloy. Whereas from the same columns in the periodic table having positive formation energies, NiPd mostly remains in disordered phases with a tendency toward short ranged clustering.

	              The NiPt alloy system has also been drawing  attention because of its magnetic \cite{beille},\cite{parra} and catalytic behaviors. The fragile and weak magnetism of Ni \cite{parra} is strongly affected by its near-neighbor configuration in the alloy.  This dependency is important particularly in disordered phases. The magnetic moment and Curie temperature vary widely with composition disordered alloys, particularly at the lower concentration of Ni. Parra and Cable \cite{parra} have experimentally studied  both local as well as average magnetic moments in Ni-Pt systems by neutron scattering experiments.  They have reported, using the Warren-Cowley short-range order parameter, that Ni-Pt alloys have a spatially homogeneous moment distribution which is in sharp contrast to Ni-Cu and Ni-Rh that exhibit ferromagnetic clustering. They reported that the magnetism of Ni-Pt alloys is sensitive to local chemical ordering. The 50-50 alloy is paramagnetic \cite{parra} when in the ordered L1$_0$ state whereas it is ferromagnetic \cite{uday1},\cite{uday2} in the disordered configuration.

                   To date there has been a series of theoretical as well as experimental investigations of the electronic and magnetic properties of Ni-Pt alloys. These include the Stoner  and Edwards-Wohlfarth  model of itinerant ferromagnetism by Alberts \etal \cite{albert} and Baillie \etal \cite{bail}, local spin density functional (LSDA) based calculations using the Korringa-Kohn-Rostocker coherent potential approximation (KKR-CPA) by Stocks and Winter \cite{kouvel},\cite{hicks}. The non self consistent, but relativistic KKR-CPA study was carried out by Staunton {\etal} \cite{stau} and a self consistent, but nonrelativistic version by Pinski {\etal} \cite{cable}. The surprising discrepancies between the various works can only be attributed to the indequate description of environmental dependence in the mean-field theory like approaches (CPA). There was a disagreement between experiment and theory in the Stoner-Edwards-Wohlfarth model because Ni-Pt is weakly ferromagnetic. In particular, the disagreement was over the composition dependence of the zero-field, zero temperature magnetization \cite{stoner}. All the previous approaches for the study of disordered alloys were based on a single site and mean field like approaches (CPA). It excludes the crucial effect of the local environment.

              We had earlier investigated successfully the electronic and magnetic properties of disordered binary NiPt alloys using a LSDA based tight-binding linear muffin tin orbitals augmented space recursion (TB-LMTO-ASR) method with scalar relativistic corrections and studied the effects of chemical disorder on the fragile magnetism of Ni \cite{durga}.  Disorder was successfully treated by the augmented space formalism (ASR) introduced by us.  The technique is capable of going beyond the single site approximation and includes local short-ranged ordering effects\footnote{For an explicit discussion on the Augmented Space Method the reader is referred to the monograph \cite{asr1}}. The  relativistic effects result in  higher local magnetic moment for Ni in NiPd as compared to that in NiPt. The local magnetic moment of Ni increases with the increasing  Pd content in NiPd alloys, whereas it decreases with the addition of Pt in NiPt \cite{singh}. In an earlier communication \cite{uday2} we have pointed out both theoretically and experimentally that NiPt alloys within (40-60)\% Ni content, is strongly depend on its environment.  ASR allows us to study effects of local environmental fluctuation with much greater accuracy than the CPA like single site and mean field approximation. Therefore, ASR is a reliable method for the studying of optical response in systems where local environmental effects are predominant.

      In this communication, we shall study the configuration averaged optical responses such as optical conductivity and complex dielectric function for ordered and disordered NiPt and Ni$_3$Pt alloys. Our approach will be equivalent to a formulation in terms of the linear response or the Kubo formalism. This formalism combines the Kubo-Greenwood \cite{kubo},\cite{greenwood} approach with a DFT based tight-binding linear muffin-tin orbitals and augmented space recursion (TB-LMTO-ASR) \cite{asr1}-\cite{asr7} techniques which explicitly incorporates the effects of disorder. We find the enhancement of optical properties due to disorder and broadening of the interband optical transitions in Ni$_{1-x}$Pt$_x$ and Ni$_{3(1-x)}$Pt$_x$ alloys with increasing Pt concentration. We shall also study the fragile magnetic effect on the optical response of Ni$_{1-x}$Pt$_x$ alloys within the composition range of (40-60)\% Ni. Our investigation shows that NiPt and Ni$_3$Pt alloys would be a promising materials for optically transparent and electrically conductive thin metalic alloy fimls (TMAFs). They  can also be  suitable  for ohmic contacts in modern complementary metal-oxide-semiconductor (CMOS) based device processing applications. 
 
\section{Methods and computational details.}

      Before we begin our analysis of the alloys, let us discuss in brief the theoretical techniques we propose to use. Our starting point will be the density functional approach (DFT) \cite{dft1}-\cite{dft4}. The effect of the many-body interactions will be taken care of in the electron density dependent Hartree and exchange-correlation potentials. The full many-body Schr\"odinger equation of the ~10$^{20}$ odd mutually interacting electrons and ion-cores are compressed by the DFT and Born-Oppenheimer approximations into the one-electron Kohn-Sham equation. In our calculations we have used both the ground-state exchange-correlation potential of von Barth and Hedin \cite{vbh} and the more recent excited state exchange potential of Harbola and Sahni \cite{hs}.

      The next step is the solution of the Kohn-Sham equations. Here we have a wide variety of well-developed methods, hence a judicious choice is essential. We note first that we shall have to study disordered solids where the ion-core potentials do not possess lattice translational symmetry and the Bloch Theorem is no longer valid. The Bloch function is no longer a solution of the Kohn-Sham equation. This is unlike in crystalline solids where the Bloch function {\sl is} a solution and a real $\vec{k}$ (Bloch wave vector) is a good quantum label for both the energies and the wavefunctions.  One way out of this dilemma is to adopt the super-cell method, where we take a random supercell of size $N$ atoms randomly distributed and then apply periodic boundary conditions. The model may describe ``local" properties reasonably well, but as long as $N$ is finite the spectrum is discrete and the spectral density a bunch of delta functions. One hopes to come out of the problem by introducing an imaginary part to $\vec{k}$, but how do we calculate it without introducing extraneous parameters is a point to ponder on. We have decided to follow Heine's advice and ``throw reciprocal space out of the window" \cite{haydock}, and adopt a purely real space approach. This is the linear combination muffin-tin orbitals (LMTO) which is a close relative of the chemists linear combination of atomic orbitals (LCAO) method.

The solutions of a single muffin-tin potential placed at $\vec{R}_n$ form the basis. Linearization 
around $E_{\nu\ell}$ gives :
$ \{ |nL\rangle \} $  with a real space representation 
 \be
  \langle\vec{r}|nL\rangle =  \phi_{nL}(\vec{r}-\vec{R}_n,E_{\nu\ell}) + \sum_{n'L'} 
h_{nL,n'L'}\ \dot{\phi}_{n'L'}(\vec{r}-\vec{R}_{n'},E_{\nu\ell}) \qquad
\ee
where $ L = (\ell,m,m_\sigma)$.
In this discrete basis the  Hamiltonian representation is a matrix of inifinite rank. To calculate
its resolvent and various correlation function the ideal technique is the Recursion method of Haydock \etal \cite{haydock}. This essentially involves a change of basis which renders the Hamiltonian matrix in 
a Jacobian, tri-diagonal, form :
\begin{eqnarray}
|1\rangle & = &  |nL\rangle \quad |2\rangle = H|1\rangle - \alpha_1 |1\rangle \nonumber\\
|m\rangle & = & H |m-1\rangle - \alpha_{m-1} |m-1\rangle - \beta^2_{m-2} |m-2\rangle\quad m \geq 2 \nonumber\\
\end{eqnarray}

       We have successfully combined three important algorithms available for electronic structure calculations : the DFT based TB-LMTO developed by Andersen and his group at Stuttgart, the Recursion Method developed by Haydock and Heine and their group at Cambridge and the Augmented Space method developed by Mookerjee and his group. This a powerful computational package which makes
 no use of the Bloch Theorem and remains firmly in real space. Therefore it can tackle disorder both chemically and structurally with equal footing. We have used two separate techniques to calculate the optical response of ordered and disordered alloys. We have used DFT based tight-binding linear muffin-tin orbitals (TB-LMTO) method to calculate the band structure and DOS of ordered alloys. Then we have used the TB-LMTO-recursion technique (RS) \cite{viswa}  to calculate the optical response of ordered alloys. To calculate the optical response for disordered alloys, we have used TB-LMTO-augmented space generalized recursion technique (ASGR) \cite{asr2}-\cite{asr7}. Augmented space formalism (ASF) \cite{asr2}-\cite{asr7} explicitly incorporates the effects of disorder. We have used the local spin-density approximation (LSDA) based tight-binding linear muffin-tin orbitals (TB-LMTO) technique using von Barth and Hedin vBH exchange-correlation functional \cite{vbh}. We introduce disorder by changing Pt and Ni concentrations in Ni-Pt system. The disordered alloys  all have cubic symmetry. The parameters for the Hamiltonian and optical current operator of binary disordered system are generated from TB-LMTO within local spin density approximation (LSDA) using Barth and Hedin exchange correlation potential. This procedure has been discussed earlier and interested reader may consult our review \cite{moshi}. We shall use the technique to calculate the matrix elements of current operator described by Hobbs {\etal} \cite{hobs}. We have calculated optical current operator and optical properties within the framework of the self-consistent linear muffin-tin orbital (TM-LMTO) band structure theory. The optical current operator is computed using output data from the TB-LMTO band structure package. We have used three shell augmented space calculation and carried out fifty  steps of recursion before our computation cut-off limit. We have used the square-root terminator for convergence of continued fraction coefficients for the current-current optical correlation function suggested by Viswanath and M\"uller \cite{viswa}.

   For uniaxial current, the optical current-current response function is given by :
 \[
C(\vec{r}-\vec{r}\ ^\prime,t) =  < J(\vec{r},t)J(\vec{r}\ ^\prime,0)> - < J(\vec{r},t)><J(\vec{r}\ ^\prime,0)>
 \]

    We calculate the configuration averaged optical current correlation function $\ll S(\omega)\gg$ which is the Laplace transform of  $\ll C(t)\gg$ , from a Kubo-Greenwood approach based TB-LMTO-ASR formalism. The imaginary part of the dielectric function is related to this correlation function through :
\[
 \epsilon_2(\omega) = 4\pi \frac{\ll\!S(\omega)\!\gg}{\omega^2}
 \]

 Here $\ll...\gg$ indicates a configurational averaged quantity of a disordered system described in augmented space formalism (ASF) \cite{asr7}. The recursive equations described in the Appendix leads to a  continued fraction expansion of the correlation function :
 
 \begin{equation}
 \ll S(z)\gg = \frac{i}{\displaystyle z-\tilde{\alpha}_1 - \frac{\displaystyle \tilde{\beta}_1^2}{\displaystyle z-\tilde{\alpha}_2-\frac{\displaystyle \tilde{\beta}_2^2}{\displaystyle z-\tilde{\alpha}_3-\frac{\displaystyle \tilde{\beta}_3^2}{\ddots}
 }}}
 \end{equation}
 
 Here, $z = \omega - i\delta^+ $. The real part of the dielectric function $\epsilon_1(\omega)$ is related to the imaginary part $\epsilon_2(\omega)$ by a Kramer's Kr\"onig relationship :

\[
 \epsilon_1(\omega) = \frac{1}{4\pi}\int_{-\infty}^\infty \ d\omega'\ \frac{\epsilon_2(\omega')}{\omega -\omega'} 
\]
And finaly, The optical conductivity is  given by :  

\[ \sigma(\omega) = \frac{\omega\epsilon_2(\omega)}{4\pi} \]

\section{Results and Discussion}
\subsection{Electronic Structure of Ni-Pt alloys}

\begin{figure}[t!]
\centering
\framebox{\includegraphics[width=3.9cm,height=3.9cm]{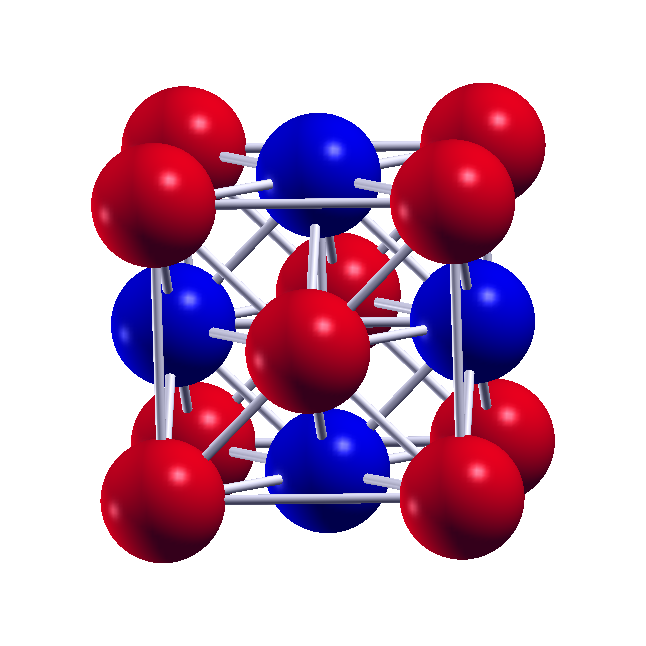}} 
\framebox{\includegraphics[width=3.9cm,height=3.9cm]{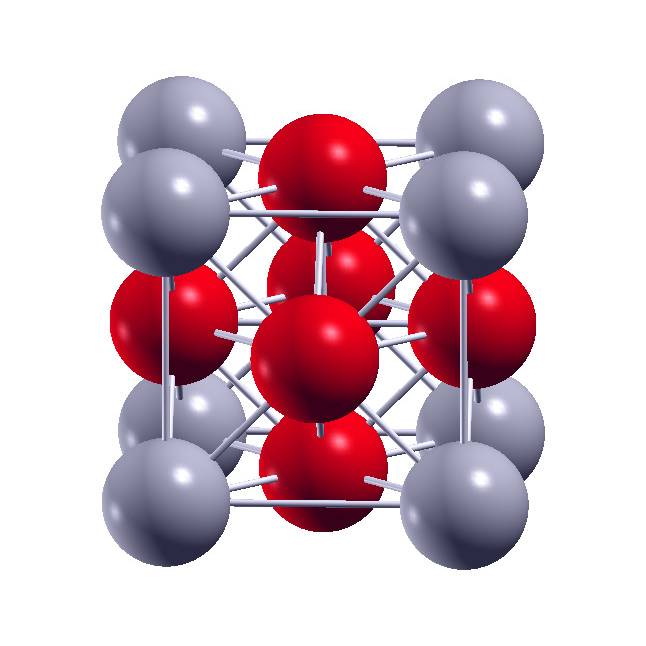}}
\caption{(Color Online) (left) Ordered L1$_0$ fcc structure of NiPt lattice. Here blue circles : Ni, red circles : Pt. (right) Ordered L1$_2$ fcc structure of Ni$_3$Pt lattice. Red circles : Ni, violet circles : Pt. 
 \label{unit}}
\end{figure}

\begin{figure}[b!]
\centering
\includegraphics[width=6.5cm,height=5.0cm]{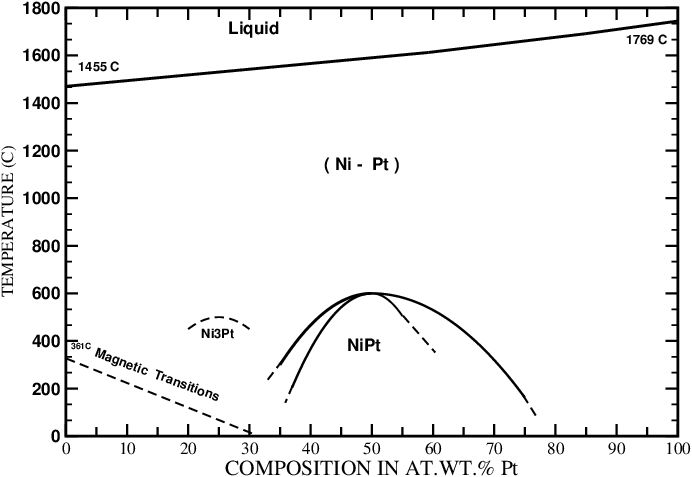}
\caption{The experimental phase diagram of Ni-Pt alloy system collected and reported by Massalski \cite{massal}.\label{pd}}
\end{figure}

   In the Ni-Pt system experiments have identified two ordered phases : NiPt and Ni$_3$Pt as shown in Fig.\ref{unit}. NiPt and Ni$_3$Pt are the two ordered ground state structures according to Massalski's Ni-Pt phase diagram \cite{massal} as shown in Fig.\ref{pd}. The ideally ordered L1$_0$ and L1$_2$ structures have the stoichiometric compositions of $50\%$ and $25\%$ of Pt. As we deviate from stoichiometry, patches of disorder appear where the excess atoms sit. When we are far from stoichiometry, the original configurations cannot be sustained any further and the disordered alloy becomes the stable phase. The Ni-Pt system is in ordered L1$_2$ (Ni$_3$Pt) phase in the composition range of  20  to 30$\%$ Pt content and  in the ordered L1$_0$ (NiPt) phase in the range 40 to  60$\%$ Pt content. On the Pt-rich side, however, only a few experimental results are available. For NiPt, a first-order order-disorder phase transition takes place at 645$^{\circ}$C, below which it crystallizes into the ordered tetragonal L1$_0$ structure. In case of Ni$_3$Pt, a cubic-to-cubic phase transition takes place at 580$^{\circ}$C into the LI$_2$ structure. At higher temperatures up to about 1420$^{\circ}$C Ni-Pt exists as a solid solution on a face-centered cubic (f.c.c.) lattice over the whole range of composition. The ordered L1$_0$ and L1$_2$ structures are the two regions of our interest in the present study. 
 \begin{table}[!h]
 \centering
 \begin{tabular}{cccl} \hline
 \hline
 Structure & Composition & Symmetry & Lattice parameters \\ \hline 
 L1$_2$ & 20-30 $\%$ Pt & Pm-3m & a=c=3.836 $\AA$ \cite{pisan}\\
 L1$_0$ & 40-60 $\%$ Pt & P4/mmm & a=3.814 $\AA$ \\
        &               &        & c=3.533 $\AA$ \cite{pearson}-\cite{pisan}\\ \hline\end{tabular}
 \caption{Experimental parameters used by us for the TB-LMTO-Recursion calculations.\label{tab1}}
 \end{table}
 In Table \ref{tab1} we show the structural parameters of the L1$_0$ and L1$_2$ Ni-Pt collected from experimental reports which we have used for our electronic structure calculations. For ordered alloys
we do not need to invoke the Augmented Space Formalism. A complete calculation can be
carried out by a combination of TB-LMTO and Recursion.  In this work we shall  present theoretical calculations for the ground state for the ordered alloys which is in agreement with previous studies. \cite{uday1}-\cite{uday2}, \cite{pearson}-\cite{dahm}.

\begin{figure}[!h]
\centering
\includegraphics[width=3.0cm,height=3.5cm,angle=270]{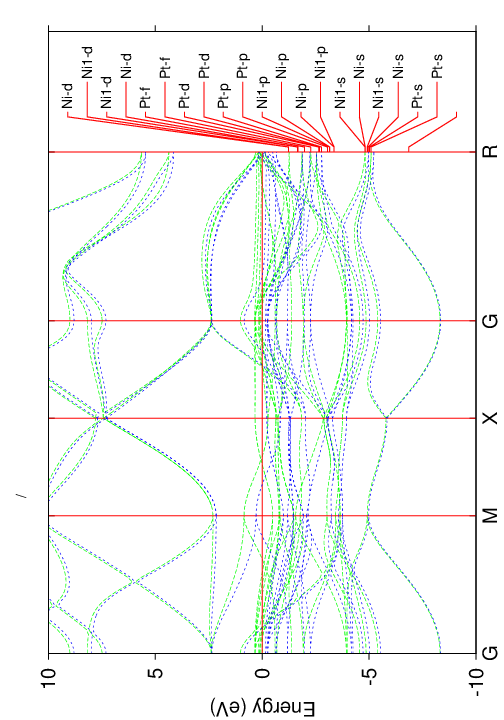}
\includegraphics[width=3.0cm,height=3.5cm,angle=270]{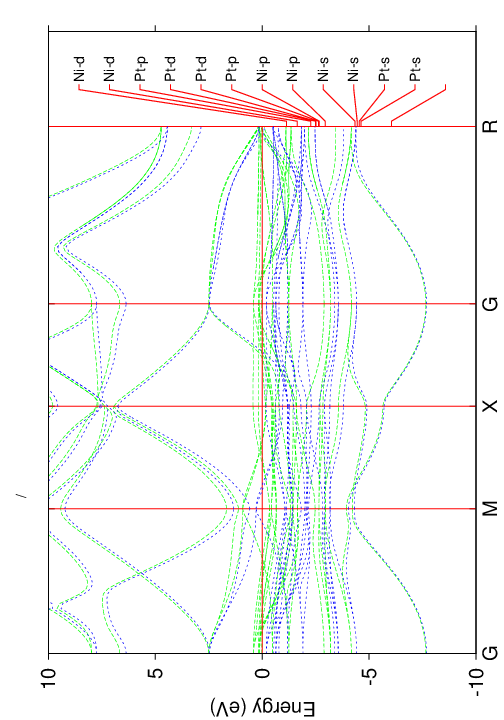}
\caption{ Band structure of ordered (left) NiPt and (right) Ni$_3$Pt along the high symmetric direction in the first brillouin zone. Fermi level is at 0 eV. The Green Blue lines indicate the up and down spin states, respectively. Bands below -5 eV and above fermi level show free electron like dispersion. This are coming from outer most s orbitals. The flat band just below the Fermi level are derived from the atomic d-orbitals. Here there is significant mixing of d and s-orbitals which are responsible for the optical interband transitions leading to peak in optical conductivity.}
\label{bands} 
\end{figure}

\begin{figure}[t!]
\centering
\includegraphics[width=4.0cm,height=3.5cm,angle=0]{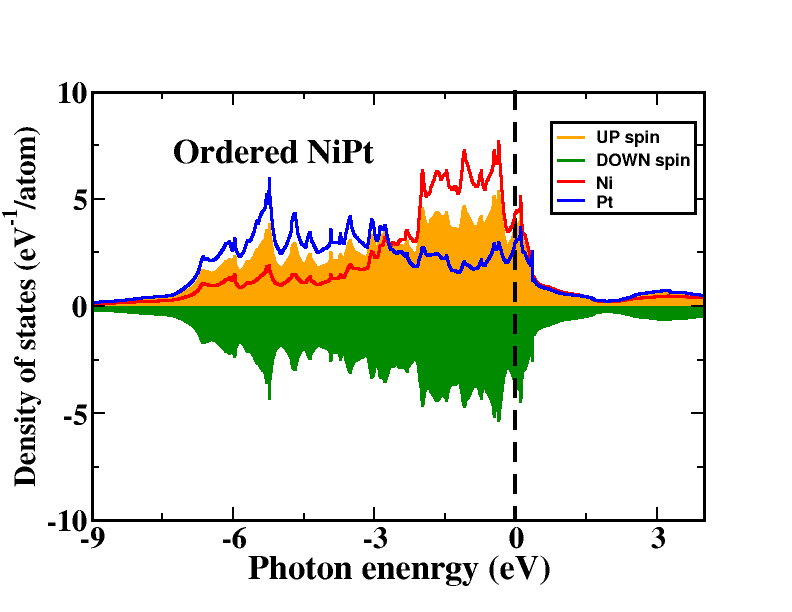}
\includegraphics[width=4.0cm,height=3.5cm,angle=0]{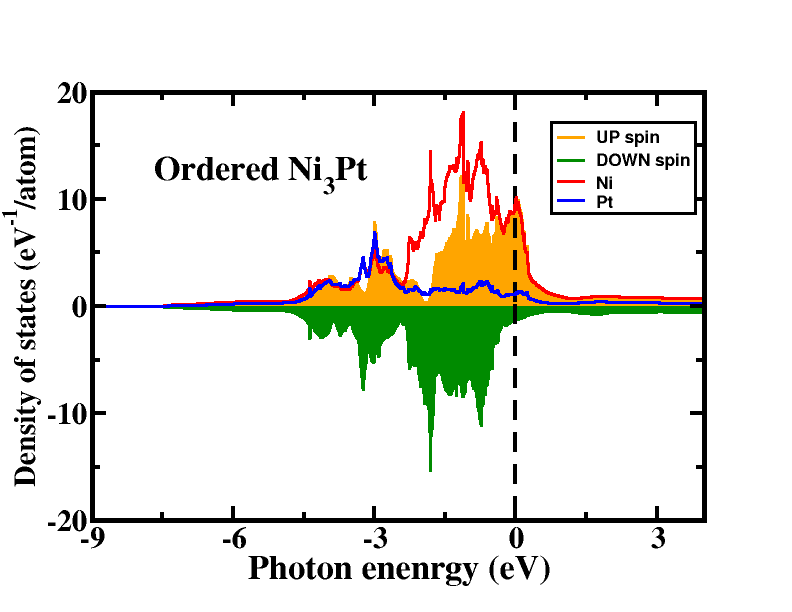}
\caption{ The total density of states of ordered (left) NiPt and (right) Ni$_3$Pt. Here the energy is measured from the Fermi energy fixed at 0 eV.}
\label{dos}
\end{figure}

    Fig.\ref{bands} shows the band structure of ordered (left) NiPt and (right) Ni$_3$Pt along the high symmetric directions in the first brillouin zone. The energy is measured from the Fermi energy  in eV. The  blue and green lines indicate the up and down spin states, respectively. Bands below the -5 eV and above the Fermi show free electron like band dispersion. The lowest s-state is approximately  at 8.35 eV for NiPt and 7.75 eV for Ni$_3$Pt with quadratic dispersion at $\Gamma$ point. These bands are from the outermost s orbitals. The flat bands straddling near the Fermi energy are derived from the atomic d-orbitals. There is significant mixing of d and s-orbitals which is responsible for optical interband transitions leading to peak in optical conductivity. Fig.\ref{dos} shows the  spin resolved density of states (DOS) calculated over the range from -20 eV to 20 eV for Ni$_3$Pt and -10 eV to 10 eV for NiPt around the Fermi level.
The d-bands of Ni and Pt are confined to a range of slightly more than 5.8 eV and all minority states lies below the Fermi level. The highest majority band state lies about 0.5 eV above the Fermi level along $\Gamma-M-X-\Gamma-R$ direction. They are the crucial bands for the interband transition in Ni-Pt alloys. The 5d electrons in Pt exhibit greater bandwidth than Ni-3d electrons. This mainly causes the broadening of UV peak in optical conductivity with increasing Pt concentration. Examining DOS near the Fermi level in lower panel of Fig.\ref{dos} reveals the another significant issue related to wider distribution of states for Pt as compared to Ni. Specifically, the distribution of states for Pt extend from -6.3 eV to 1 eV (approximately) whereas most Ni states are confined within the range of -2.9 eV to 0.8 eV (approximately) in NiPt alloy. In Ni$_3$Pt alloy Ni has higher density of states than Pt. Calculations for the electronic structure of both NiPt and Ni$_3$Pt are done within the local spin dependent density functional approximation (LSDA) at the energy minimum value of lattice parameter. We now calculate the spin-projected density of states and
\[ m \mbox{\rm (spontaneous magnetization) } = \ \int_{-\infty}^{E_F} dE \{n_\uparrow(E)-n_\downarrow(E)\}\]              
$m$ is a measure of zero temperature magnetic moment. From Fig.\ref{dos}, it appears that for Ni$_3$Pt $\Delta n(E) = n_\uparrow(E)-n_\downarrow(E) \neq 0$ while for NiPt $\Delta n(E) = 0$. This is an indication that Ni$_3$Pt may have a low temperature magnetic state while ordered NiPt has a non-magnetic one.

\subsection{Optical response and effect of disorder. }

\subsubsection{Optical response of ordered alloys. }

\begin{figure}[b!] 
\centering
\includegraphics[width=4.0cm,height=4.3cm,angle=0]{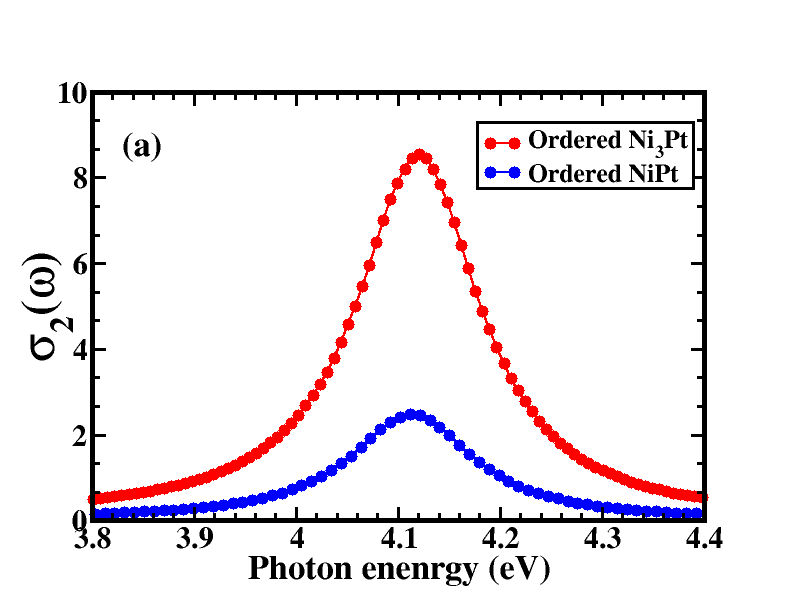}
\includegraphics[width=4.0cm,height=4.3cm,angle=0]{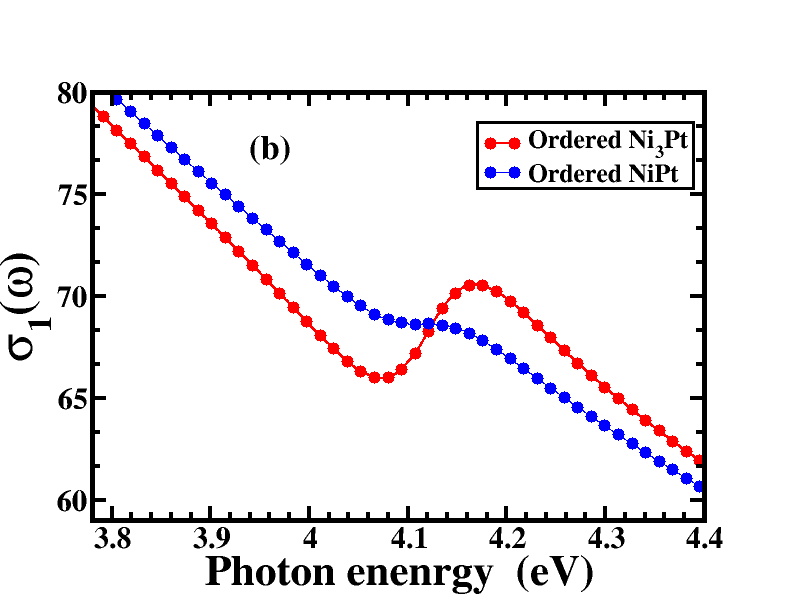}\\ 
\includegraphics[width=4.0cm,height=4.3cm,angle=0]{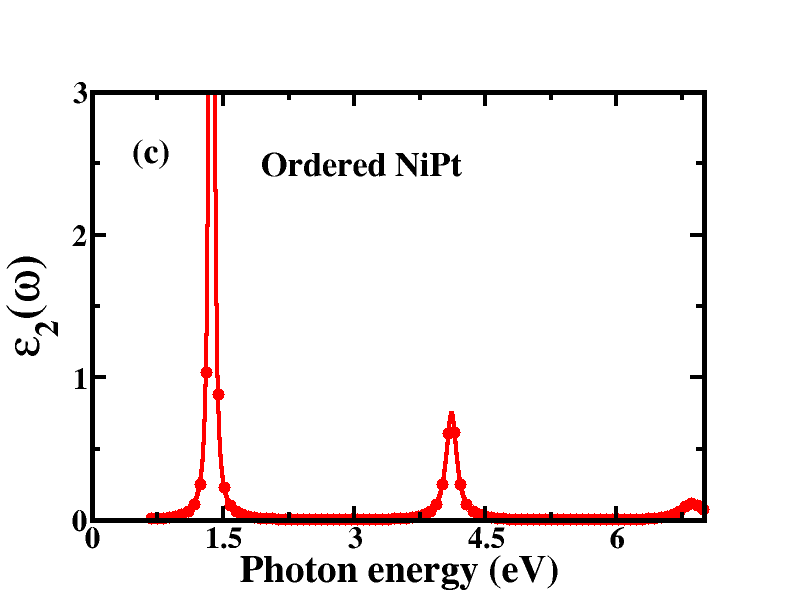}
\includegraphics[width=4.0cm,height=4.3cm,angle=0]{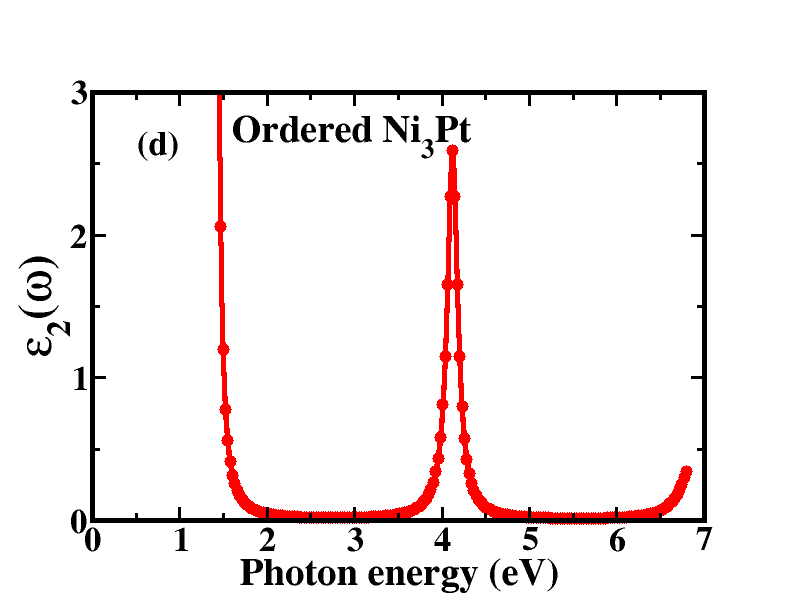}
\caption{(Color Online) (a) The imaginary part $\sigma_2(\omega)$ and (b) the real part  $\sigma_1(\omega)$ of the optical conductivity  $\sigma(\omega)$ and the imaginary part of dielectric function $\epsilon_2(\omega)$ for ordered (c) NiPt and (d) Ni$_3$Pt. Both the ordered structure have the optical conductivity transition peak at 4.12 eV.
\label{opt1}}
\end{figure}

        Figure \ref{opt1} shows the real and imaginary part of optical conductivity of ordered NiPt and Ni$_3$Pt alloys. Ordered NiPt has 50\% Pt concentration and that of ordered Ni$_3$Pt it is 25\%. A sharp absorption peak is observed at 4.12 eV for both the ordered NiPt and Ni$_3$Pt alloys due to interband optical transition. In a  recent experiment, Abdallah $\etal$ \cite{abda} have reported similar transition peak at 4.5 eV in 10 nm thick annealed NiPt film using spectroscopic ellipsometry and Drude-Lorentz oscillator fitting. The optical properties are strongly dependent on the environment, particularly at the larger Pt concentrations and depends on the exact form of the exchange-correlation functional in TB-LMTO technique. Our study is based under this consideration. From band structure calculation as shown in Fig.\ref{bands}, we therefore expect that optical transition at 4.12 eV is due to transition from the occupied d bands below the Fermi level to the unoccupied free electron like sp bands above the Fermi energy. It is at symmetric M point in the first brillouin zone for both alloys. The optical peak positions for ordered alloys can be deduced from the band structure. Since the disorder induced compositional fluctuations are time independent (annealed disorder) the transitions are equi-energetic. This is the vertical transitions between the occupied states in the vicinity of the Fermi energy and any of the lower unoccupied states. With disorder the bands broaden and the peaks get width. It is comparable to the peak in spectra of imaginary part of the dielectric function ($\epsilon_2$) clearly depicted in lower panel of Fig.\ref{opt1}. In NiPt alloy a strong divergent peak is observed at 1.4 eV near-IR photon energy as shown in Fig \ref{opt1}. This divergent peak is due to the transitions within 3d bands of majority carriers near the Fermi edge. Because of the strong divergence of $\epsilon_2$ at low energies the srructure at 1.4 eV is not visible in the $\epsilon_2$ data of Ni$_3$Pt alloy. Our results are in excellent agreement with those tabulated by Lynch and Hunter group \cite{lynch} and Johnson and Christy group \cite{john} as displayed in Table \ref{tab}. Lynch and Hunter found the conductivity transition peak at 1.5 eV (1.4 eV in our work) in low temparature region and the peak at 4.05 eV (4.12 eV in our work) is strong at room temparature as precribed by Johnson and Christy. Here we are interested on the high photon energy spectrum (UV). The UV peak becomes sharper dip in ordered Ni$_3$Pt than ordered NiPt because of lower Pt concentarion. The coupling between 3d and 5d transition metals plays a significant role in Ni-Pt optical absorption. The 3d-5d bonding in Ni-Pt system is responsible for broadening of the UV optical peak. In Figure \ref{opt1}, the opticla UV peak occurs at 4.12 eV due to inter d-band transition for both ordered NiPt and Ni$_3$Pt. 5d electrons in Pt have higher band width as compared to the 3d electrons in Ni. As d-bands has a strong impact on optical interband transition of Ni-Pt alloys, the UV peak broadens with increasing Pt concentration. Drude like behaviour occurs for ordered Ni$_3$Pt alloy only below 1.5 eV. We will show later the effect of disorder by deviating from the stoichiometric compositions by adding Pt concentration in Ni-Pt alloys. In absence of disorder, the correlation function ${\rm S}(\omega)$ is a bunch of Dirac delta functions. It widens to a Lorenzian with increasing disorder strength. Abdallah $\etal$ have done their experiment on both ordered Ni$_{1-x}$Pt$_x$ alloys with x= 0, 0.1, 0.25 and disordered Ni$_{1-x}$Pt$_x$ alloys with x= 0.1, 0.15, 0.2, 0.25 respectively. They obtained the optical conductivity peak of ordered Ni$_{1-x}$Pt$_x$ alloy at 4.5 eV for all compositions with the broadening of UV peak with increasing Pt content. At x=0 content (ordered Ni), they have compared their results with the experimental observation of Lynch $\etal$ \cite{lynch} at 4.05 eV and Johnson $\etal$ \cite{john} at 4.1 eV as displayed in Table \ref{tab}. Abdallah $\etal$ have done spectroscopic ellipsometry experiment on 10 nm Ni$_{1-x}$Pt$_x$ with different concentration (0.0$\leq$ x $\leq$0.25) while our calculations are on bulk Ni$_{50}$Pt$_{50}$ and Ni$_{75}$Pt$_{25}$. The interactions between Si substrates and samples are taken into consuderation in ellipsometry measurements of Abdallah $\etal$. Therefore, The agreement of our theoretical calculation on ordered Ni$_{50}$Pt$_{50}$ and Ni$_{75}$Pt$_{25}$ are quite well with the ellipsometry results as well as other experimental observation.

\begin{table}[h!]  
\centering
\begin{tabular}{lll}\hline\hline
 &  &  \\
\multicolumn{3}{c}{Peak position in optical conductivity}\\ 
 &  &  \\\hline
  &  &  \\
 \multicolumn{3}{c}{Comparison between Experiment and Theory}\\ 
  &  &  \\\hline
   &  &  \\
Alloy Structure & Experimental Results & Theory \\ 
 &  &  (Our Work) \\
  &  &  \\\hline
   &  &  \\
\multicolumn{3}{c}{Ordered Alloys} \\ 
 &  &  \\\hline
  & &  \\
 Ni  & 4.1 eV (Johnson $\etal$ \cite{john}) &     \\
  &  &     \\
 Ni  & 4.05 eV (Lynch $\etal$ \cite{lynch}) &   \\
 &  &    \\
 NiPt & 4.5 eV (Abdallah $\etal$ \cite{abda}) & 4.12 eV  \\
 & &  \\
  & &  \\
Ni$_3$Pt  & &  \\
 & $-$ &  4.12 eV  \\
  & &  \\
 & &  \\ \hline
 & &  \\
\multicolumn{3}{c}{Disordered Alloys} \\ 
  &  &  \\\hline
      &  &  \\
& &  \\
Ni$_{1-x}$Pt$_x$  & 5.3 eV (Johnson $\etal$ \cite{john}) & \\
(0.1$\leq$ x $\leq$0.25) & 5.0 eV (Abdallah $\etal$ \cite{abda}) &  5.5 eV \\
  & &  \\
 & &  \\
Ni$_{1-x}$Pt$_x$  &  $-$ & 3.93 eV \\
(0.4$\leq$ x $\leq$0.6) &  & \\
& &  \\
& &  \\
Ni$_{3(1-x)/3}$Pt$_x$   &  $-$ & 5.6 eV \\
(0.1$\leq$ x $\leq$0.3) &  & \\
& &  \\
\hline \hline
\end{tabular}
\caption{Optical conductivity transition peak position for ordered and disordered NiPt and Ni$_3$Pt alloys. Number in the square brackets represents the reference numbers. Our calculations were all on  bulk systems. Experiments are also carried out on thin films.
\label{tab}}
\end{table}

\subsubsection{Optical response of disordered alloys.}

\begin{figure}[b!]
\centering
\includegraphics[width=4.0cm,height=4.3cm,angle=0]{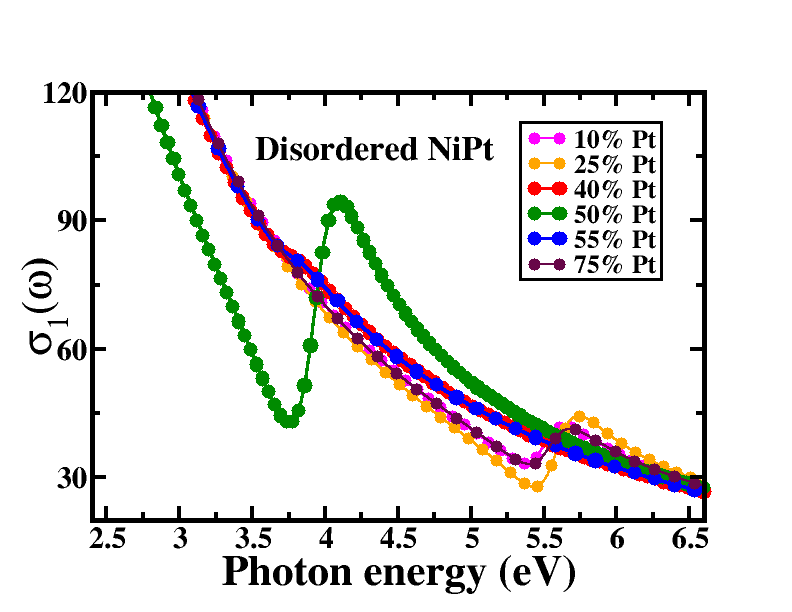}
\includegraphics[width=4.0cm,height=4.3cm,angle=0]{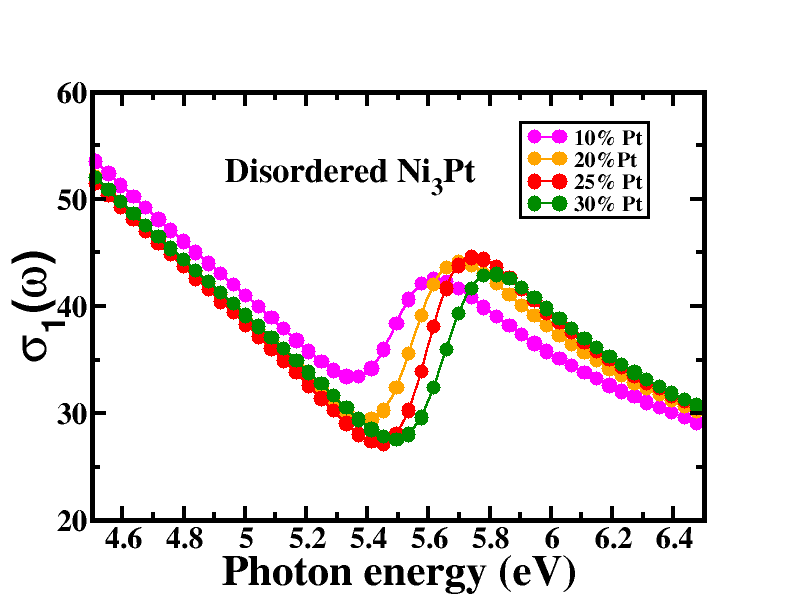}\\ 
\includegraphics[width=4.0cm,height=4.3cm,angle=0]{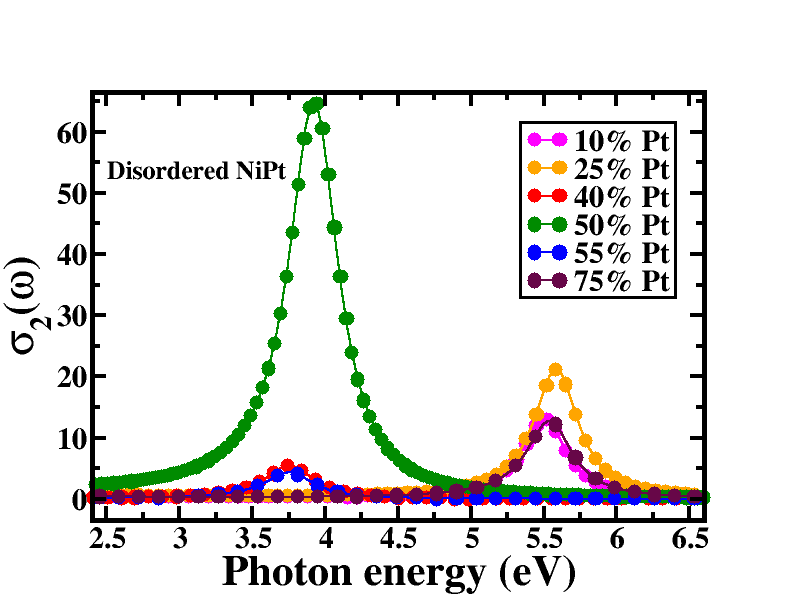}
\includegraphics[width=4.0cm,height=4.3cm,angle=0]{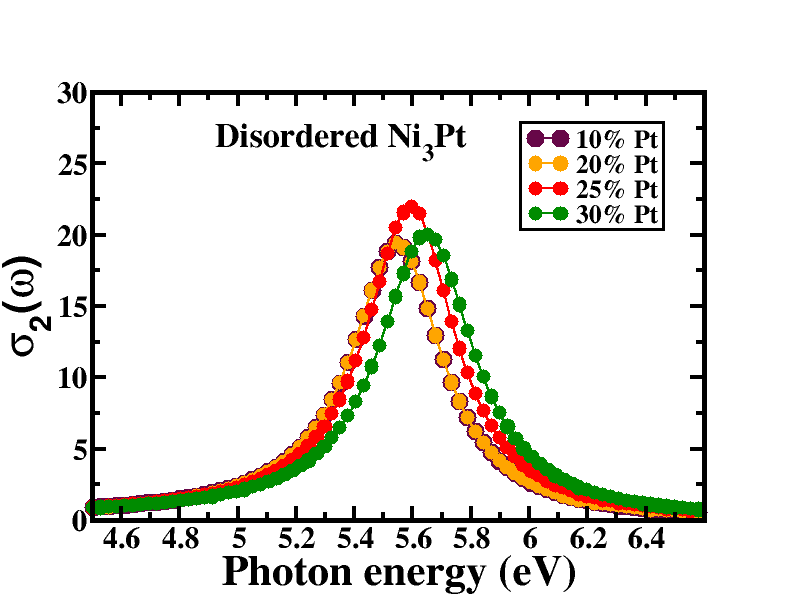}
\caption{(Color Online)  The real $\sigma_1(\omega)$ and imaginary part $\sigma_2(\omega)$ of optical conductivity  $\sigma(\omega)$ for (left panel) disordered Ni$_{1-x}$Pt$_x$ and (right panel) disordered Ni$_{3(1-x)/3}$Pt$_x$. For disordered NiPt we have used x=0.1, 0.25, 0.4, 0.5 0.55, 0.75 ie 10\%, 25\%, 40\%, 50\%, 55\%, 75\% and that for disordered Ni$_3$Pt we have used 10\%, 20\%, 25\%, 30\% Pt concentration respectively. 
\label{opt2}}
\end{figure}

\begin{figure}
\centering
\includegraphics[width=4.0cm,height=4.3cm,angle=0]{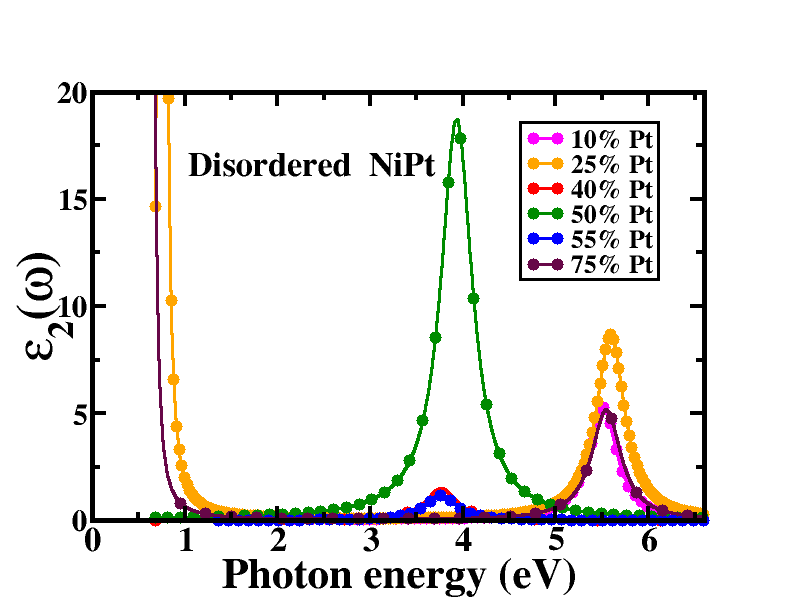}
\includegraphics[width=4.0cm,height=4.3cm,angle=0]{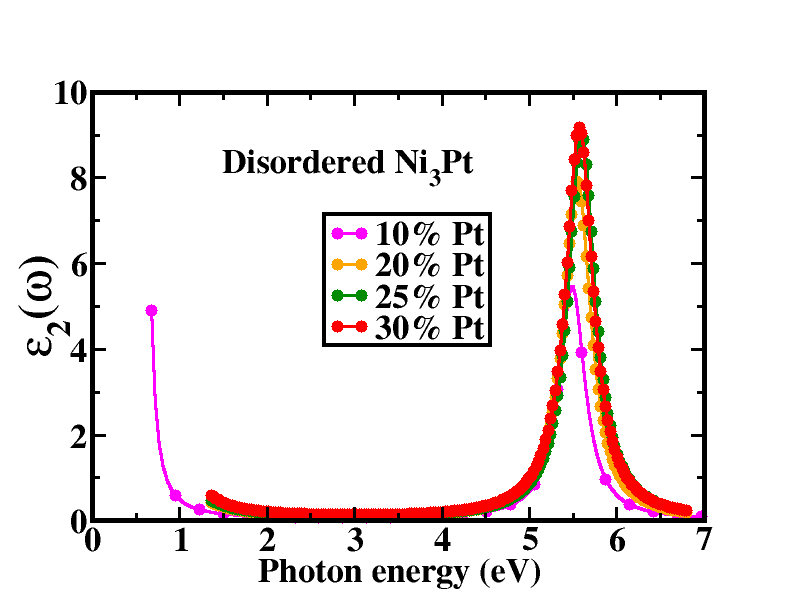}
\caption{(Color Online)  The imaginary part of dielectric functions  $\epsilon_2(\omega)$ for disordered (left) Ni$_{1-x}$Pt$_x$ and (right) Ni$_{3(1-x)/3}$Pt$_x$. Disordered Ni$_{1-x}$Pt$_x$ shows a Drude-like behaviour at low photon energies, whereas ordered NiPt does not show this.
\label{diel}}
\end{figure}

       The real and imaginary part of optical conductivity of disordered L1$_0$ Ni$_{1-x}$Pt$_x$ and L1$_2$ Ni$_{3(1-x)/3}$Pt$_x$ with increasing Pt content (x) are shown in Fig.\ref{opt2}. We take the disordered L1$_0$ alloys in the composition range of 0.1$\leq$ x $\leq$0.75 and L1$_2$  in 0.1$\leq$ x $\leq$0.3 respectively \cite{massal}. The optical peak becomes more pronounced and less sharp due to disorder induced brodening in the correlation function $\ll {\rm S}(\omega)\gg$. These peaks become resonances from ordered to disordered structure and exhibit noticeable broadening with increasing disorder strength (x). The  absorption peak in optical conductivity is observed at 3.93 eV for disordered Ni$_{0.5}$Pt$_{0.5}$ alloy. We obtained the peak near 3.93 eV for Ni$_{1-x}$Pt$_x$ with x=0.4, 0.5, 0.55. The peak shifts to near 5.6 eV for Ni$_{1-x}$Pt$_x$ with x=0.1, 0.25 and 0.75. The peak is near 3.93 eV with (40-55)$\%$ Pt content. It remains in NiPt phase within that Pt range. Disordered Ni$_{3(1-x)/3}$Pt$_x$ shows the  transition peak at 5.6 eV with (20 - 30)\% Pt content. It remains in Ni$_{3}$Pt within that Pt limit. For NiPt the peak shifts to a higher energy value near 5.6 eV for other Pt concentration outside  (40 - 55)\% Pt range. This indicates a phase transition from  the L1$_0$ to L1$2$ structure which is consistent with Massalski's Ni-Pt phase diagram \cite{massal}. This is clear from optical conductivity peak of NiPt. Ni$_{1-x}$Pt$_x$ has transition peak at 3.93 eV within x = (0.4 - 0.55) range and at 5.6 eV for x = 0.25 value. It indicates that if we change Pt from 25\% to 40\% then phase transition takes place from  Ni$_3$Pt to NiPt. Optical peak  is obtained for disordered Ni$_{0.9}$Pt$_{0.1}$ and Ni$_{0.25}$Pt$_{0.75}$ at 5.6 eV. The imaginary part of the dielectric constant ($\epsilon_2$) for Ni$_{1-x}$Pt$_x$ and Ni$_{3(1-x)/3}$Pt$_x$ are shown in Fig.\ref{diel}. The optical peak position in $\epsilon_2$ reflects the optical conductivity like nature. Abdallah $\etal$ have obtained similar peak at 5.0 eV for disordered Ni$_{1-x}$Pt$_x$ alloys (0.1$\leq$ x $\leq$0.25) from spectroscopic ellipsometric technique as tabulated in Table \ref{tab}. Ellipsometry results also include substrate-sample interaction. Johnson and Christy \cite{john} obtained similar transition peak at 5.3 eV (5.6 eV in our work) is due to transitions from lowest d bands to the free electron like bands.

	The coupling between 3d and 5d transition metals play a significant role in the optical response of Ni-Pt systems. The 3d-5d bonding is responsible for the broadening of UV optical peak. In disordered NiPt alloys, the d band of the constituents are very different. 5d electrons in Pt have higher band width as compared to the 3d electrons in Ni. Coupling between Ni and Pt leads to strong off-diagonal disorder and this leads to strong disorder scattering which has a large impact on the optical response. Therefore, with the increase of Pt concentration the off-diagonal disorder increases leading to the broadening of optical conductivity UV peak. This causes the enhancement of optical inter d-band transition with increasing Pt content. This interpretation is  supported by the band structures  for NiPt and Ni$_3$Pt alloys are shown in Fig.\ref{bands}. The Ni-3d bands offer more states and hence more transitions to occur. This keeps the UV peak magnitude fixed but brodens the transition peak. The imaginary part of the dielectric function $\epsilon_2$ for both disordered Ni$_{1-x}$Pt$_x$ and Ni$_{3(1-x)/3}$Pt$_x$ alloys with increasing Pt content (x) are shown in Fig.\ref{diel}. Photon energy greater than 2 eV for Ni$_{1-x}$Pt$_x$ and 3.5 eV for Ni$_{3(1-x)/3}$Pt$_x$ shows optical transition due to d-bands. Drude like behaviour therefore occurs below 1.5 eV for both the structures. But Drude contribution to the dielectric function of Ni$_{1-x}$Pt$_x$ alloys do not change significantly with increasing Pt content which is also in agreement with the  Abdallah $\etal$ observation. The Figure shows a small correction due to disorder effect on the optical current terms.

\begin{figure*}[t!]
\vskip -1cm
\centering
\includegraphics[width=9.0cm,height=5.0cm,angle=0]{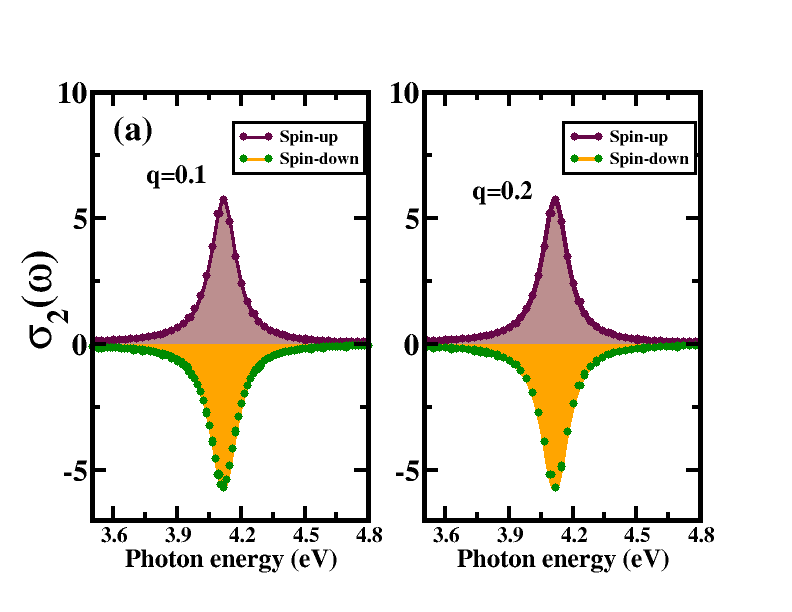}
\includegraphics[width=9.0cm,height=5.0cm,angle=0]{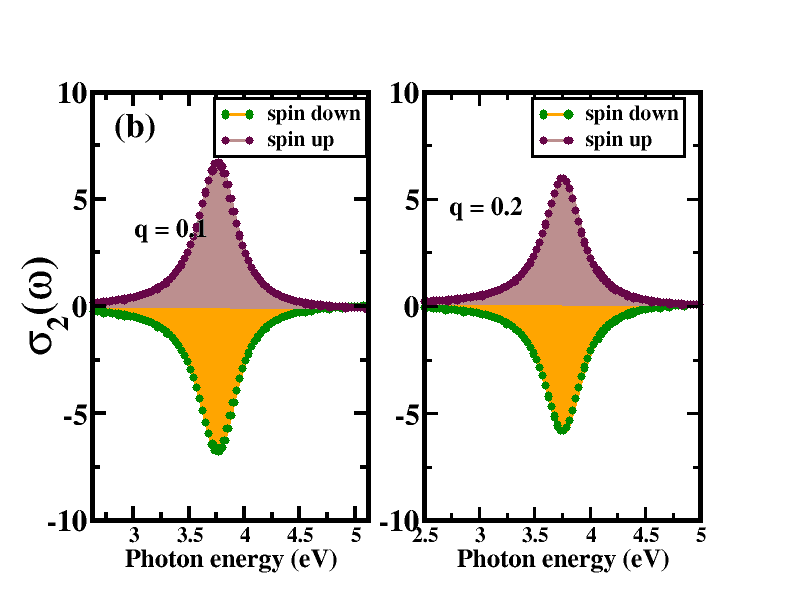}\\
\includegraphics[width=9.0cm,height=5.0cm,angle=0]{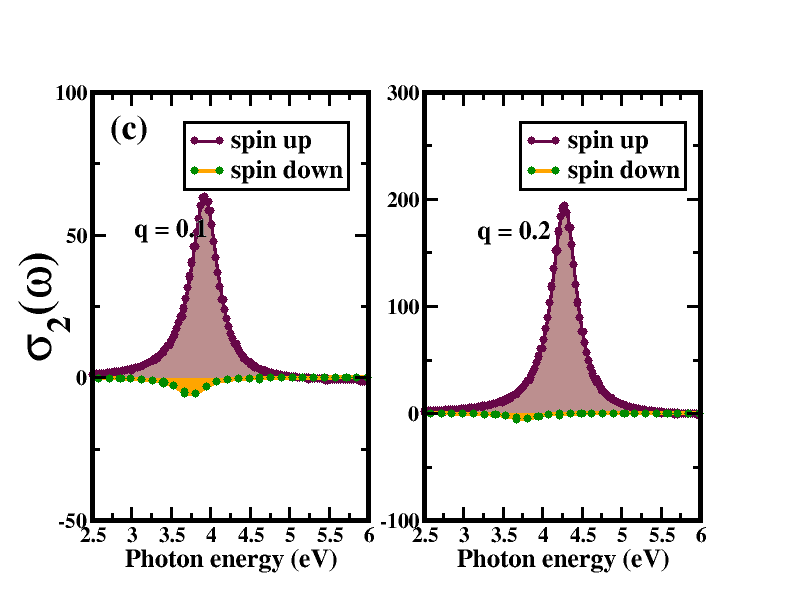}
\includegraphics[width=9.0cm,height=5.0cm,angle=0]{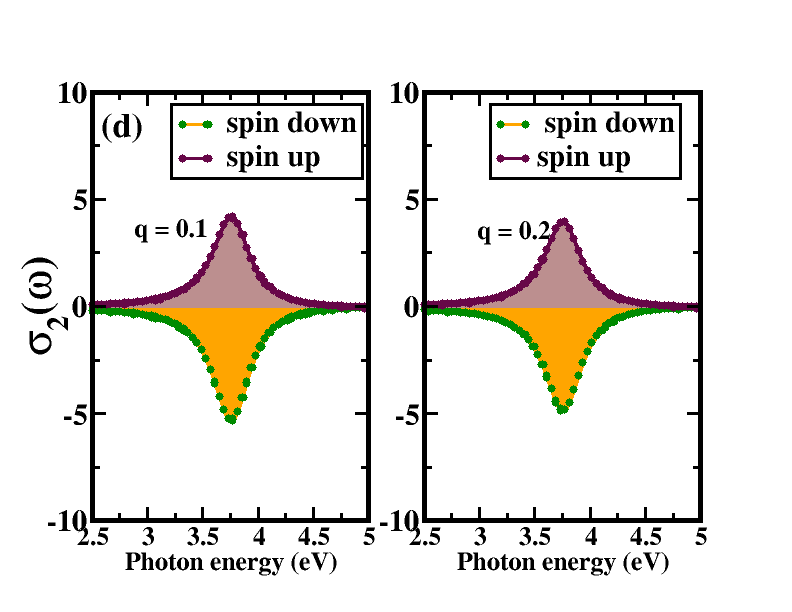}
\caption{(Color Online) Spin resolved optical response of (a) ordered NiPt and disordered  Ni$_{1-x}$Pt$_x$ (0.4$<$ x $<$0.6) ie  (b) 40\%, (c) 50\% and (d) 55\% Pt content in disordered binary alloys. Here $\vec{q}$ is the reciprocal lattice vector defined as ${\AA}^{-1}$.
\label{opt3}}
\end{figure*}

		The 3d-5d bonding is important in the optical response of NiPt and Ni$_{3}$Pt alloys. In our previous communication \cite{durga}, we studied the formation energies of the L1$_0$ structure of {NiPt and showed the effect of relativistic corrections (including mass–velocity and Darwin corrections but without spin–orbit couplings) on it. Relativistic effect significantly reduces the elastic energy of formation of 3d-5d NiPt alloys from 31.48 to 22.22 mRyd/atom \cite{durga} ( from 40.38 to 29.74 in Wang and Zunger's calculation \cite{wang} ). In addition to, relativistic corrections enhance the chemical energy of formation from −27.04 to −31.39 mRyd/atom. The contraction of the s wavefunction of Pt and the subsequent increase of s-d hybridization must be responsible for the reduction of the size mismatch, and hence reduces the strain in NiPt alloys giving rise to the stable structures. The relativistic lowering of the s potential from from -347.2 to -349.4 mRyd/atom for Ni and -283.8 to -524.0 mRyd/atom \cite{durga} for Pt causes the s-wavefunction to contract, leading to a contraction of the lattice. The energy difference of 3d-5d bonds is -138.6 mRyd/atom in the non-relativistic limit while it reduces to -114.3 mRyd/atom in a relativistic limit. The relativistic upshift of 5d-Pt band brings it closer to 3d-Ni band. This results the electron transfer from antibonding 5d bands to bonding 6s,p bands and enhances the 3d-5d bonding with adding Pt content. This increasing 3d-5d interaction and  s-d hybridization with the addition of Pt content is responsible forthe brodening of optical conductivity peak in ordered NiPt alloys than Ni$_{3}$Pt which is clearly depicted in Fig.\ref{opt1}. These effects can be appreciated by inspecting the calculated atom-projected density of states as shown in Fig\ref{dos}. We can see that the Ni and Pt states are closer to each other in NiPt, rather than in Ni$_{3}$Pt alloys. The wider Ni-3d and Pt-5d spectra  overlap more in NiPt than in Ni$_{3}$Pt. This is a signature of higher overlapping of 3d-5d bonds and hence the more negative formation energy in NiPt than in Ni$_{3}$Pt which explains the broadening of NiPt optical peak over Ni$_{3}$Pt. We conclude that the 3d-5d interaction plays a key role in enhancement of optical conductivity in Ni-Pt system with increasing Pt concentration. Disordered NiPt alloys has a optical transition peak at 3.93 eV whereas for disordered Ni$_3$Pt it is at 5.6 eV. The difference between the majority and minority spin for d band centres increases with increasing Ni content. The exchange-induced splitting for the d band is higher in Ni$_{3}$Pt than NiPt which shifts the optical conductivity peak by 1.67 eV to ultraviolet range at 5.6 eV.

\subsection{Effect of local magnetization on optical response.}

   The spin resolved, imaginary part of optical conductivity for ordered NiPt as well as disordered Ni$_{1-x}$Pt$_x$ alloys in the composition range of 0.4$\leq$ x $\leq$0.6 are shown in Fig.\ref{opt3}. We have carried out a theoretical analysis of the optical conductivity for different compositions, within ASR based density functional calculations using the local spin-density approximations. This pictures revel the effect of magnetism on the optical behaviour of NiPt alloy. For ordered NiPt, the spin projected density of states have been calculated in a LSDA (local spin dependent density functional approximation) formalism. The results shown in Fig.\ref{dos} show that for the lowest energy state, the density of up and down spin states are identical. We therefore expect a non-magnetic ground state for ordered NiPt. For disordered Ni$_{0.5}$Pt$_{0.5}$, the up-spin optical conductivity peak is higher than the down-spin one. This indicates its ferromagnetic  nature. The alloy with 40\% Pt  in the disordered structure, is non-magnetic in nature. There is a magnetic transition to a ferromagnetic state at 50\% Pt content. Disordered Ni$_{0.45}$Pt$_{0.55}$ is weakly ferromagnetic. This decrease is sharp as Pt concentration increases. In an earlier work \cite{uday2} we had carried out  both experimental and theoretical study of magnetism in disordered NiPt alloys. The conclusion stated there was that the fragile moment of Ni is strongly affected by local chemical composition. We therefore need to carry out calculations with short-ranged order included. This is possible through our ASR technique \cite{sro1},\cite{sro2}. Disorder has a crucial effect on the magnetization and hence on optical response in the (40-60)\% Ni compositions of Ni-Pt alloys. The local magnetic moment of Ni decreases with the addition of Pt content in disordered NiPt alloys. In the absence of local environmental effects, increase of Pt  should lead to an increase in the local Ni moment, since isolated atoms of Ni in Pt become more probable due to Ni clustering. Due to fragile magnetic behaviour of Ni, disordered NiPt is strongly ferromagnetic when at least 50\% of nearest-neighbour sites are occupied by Ni. This leads to narrowing of the local density of states on Ni and hence higher moment. This picture changes if there is short-ranged ordering rather than clustering. This decreases the local moment of Ni in alloy with 55\% concentration of Pt. This is due to short range ordering rather than clustering of Ni. Early experiments, \cite{uday1}-\cite{uday2} done by our group confirmed the theoretical predictions in disordered phase of NiPt alloys. Earlier total energy calculations as a function of short range order confirmed the ordering tendency in these systems. The effect of short-range ordering on magnetism plays the significant role in the optical response of disordered NiPt alloys.

\section{Conclusions}

	In conclusion, We have studied the optical response of both ordered and disordered NiPt and Ni$_3$Pt alloys using Kubo-Greenwood formalism and DFT based TB-LMTO-ASR technique. Our approach includes the correct trend in both optical and magnetic responses with increasing Pt concentration which fails to describe by single-site mean field coherent potential approximation (CPA) like approach. Disordered Ni$_{3(1-x)/3}$Pt$_x$ (0.1$\leq$ x $\leq$0.3) has a optical conductivity transition peak at 5.6 eV whereas disordered Ni$_{1-x}$Pt$_x$ (0.4$\leq$ x $\leq$0.6) has transition peak at 3.93 eV. However, it shows a phase transition clearly from Ni$_3$Pt to NiPt if we increase Pt content. We strongly believe that our study indentify the two distinct phases of Ni-Pt system according to Massalski's Ni-Pt phase diagram through their optical response.  NiPt alloy shows a significant broadening of the UV peak with increasing Pt content as Pt-5d states has higher bandwidth than Ni-3d} states. We conclude that the 3d-5d interaction plays a key role in enhancement of optical conductivity in Ni-Pt system with increasing Pt concentration. We have shown that ordered Ni$_{50}$Pt$_{50}$  alloy is non-magnetic,  whereas disordered Ni$_{50}$Pt$_{50}$ is ferromagnetic in nature. Our present observations on magnetism in Ni-Pt support the previous reports by Parra and Cable\cite{parra} and Kumar {\it et.al}\cite{uday1}-\cite{uday2}. Fragile magnetic behaviour affects the spin resolved optical response of Ni$_{1-x}$Pt$_x$ (0.4$<$ x $<$0.6). The strong environmental dependence of Ni$_{1-x}$Pt$_x$ is the cause of its fragile concentration dependency which changes the electronic structure as well as optical and magnetic response of disordered Ni-Pt systems. Our investigations have shown that not only magnetic properties but also optical properties can be tuned using chemical disorder. We anticipate that the theoretical results here will provide a useful reference and  make Ni-Pt alloys  an important material for opto-industrial metrology and semiconductor industry.

\begin{widetext}

\section*{Appendix}

          The calculations begin with a thorough electronic structure calculation using the tight-binding linear muffin-tin orbitals method. The technique, like many others, is based on the density functional theory, where the energetics depends entirely upon the charge and spin densities. However, the current operator which is characteristic of an excited electron : excited optically, electronically or magnetically, is described by transition probabilities which depend upon the wavefunction. The wavefunctions are expressed as linear combinations of the linearized basis functions of the LMTO. The wavefunction representation in the LMTO basis is :

\be
 \Psi(\vec{r})  =  \sum_{nL} c_{nL} \left[\phi_{nL}(\vec{r}-\vec{R}_n,E_{\nu\ell})
  + \sum_{nL,n'L'} 
h_{nL,n'L'}\ {\phi}'_{n'L'}(\vec{r}-\vec{R}_n,E_{\nu\ell}) \right]
\ee
where $\phi'_{nL}  =   \partial \phi(\vec{r},E)/\partial E $

The coefficients $c_{nL}$ are available from our TB-LMTO secular equation. To obtain the currents we
follow the procedure of Hobbs {\etal} \cite{hobs}.
 The optical current operator is given by :

\[ J^\mu_{RL,R'L'}\ =\ \left[ V^{(1),\mu}_{RL,RL} \delta_{RR'} + \sum_{L"} \left\{ V^{(2),\mu}_{R,R"}h_{RL",R'L'} + h_{RL,R"L"} V^{(3),\mu}_{R"L",R'L'}\right\} + \sum_{RL}\sum_{R"L"} V^{(4),\mu}_{R"L",R"',L"'}h_{R"'L"',R'L'} \right] \]

where,

\begin{eqnarray*}
 V^{(1),\mu}_{RL,RL'} & = & \int_{r<s_R} d^3\vec{r}\ \phi^{*}_{RL'}(\vec{r},E_{\nu\ell}) (-i\nabla^\mu) 
                 \phi_{RL}(\vec{r},E_{\nu\ell '}) \hskip 2cm
 V^{(2),\mu}_{RL,RL'} = \int_{r<s_R} d^3\vec{r}\ \widehat{\phi}^{*}_{RL'}(\vec{r},E_{\nu\ell}) (-i\nabla^\mu) 
                 \phi_{RL}(\vec{r},E_{\nu\ell '}) 
\\
V^{(3),\mu}_{RL,RL'} & = & \int_{r<s_R} d^3\vec{r}\ \phi^{*}_{RL'}(\vec{r},E_{\nu\ell}) (-i\nabla^\mu) 
            \widehat{\phi}_{RL}(\vec{r},E_{\nu\ell '}) \hskip 2cm
 V^{(2),\mu}_{RL,RL'} = \int_{r<s_R} d^3\vec{r}\ \widehat{\phi}^{*}_{RL'}(\vec{r},E_{\nu\ell}) (-i\nabla^\mu) 
              \mbox{\.{$\phi$}}_{RL}(\vec{r},E_{\nu\ell '}) 
\end{eqnarray*}

We have followed the prescription of Hobbs \etal\cite{hobs} to evaluate the integrals above. For details we again refer to reader to the above reference. We
should note that we have so far not introduced the idea of reciprocal space. We expect our methodology
to be applicable to situations where Bloch Theorem fails and it is not correct
 to label states and energies with a real $\vec{k}$. It now remains for us to incorporate disorder.
 This we shall do through the Augmented Space Formalism (ASF) introduced by one of us \cite{asr1}-\cite{asr6} and discussed extensively in a series of publications \cite{asr1} - \cite{asr6}.  The basic idea has
 been adopted from measurement theory. If the Hamiltonian has a random parameter $n_i$ then we can associate
 with it an operator $N_i$ such that any measurement of that parameter will lead to one of the operator's eigenvalues $\lambda^i_m$. The eigenvectors $|\lambda^i_m\rangle$ are the `configuration states' and they span the `configuration space' $\phi_i$. The spectral density of $N_i$ is the probability distribution of the random variables. The Augmented Space Theorem \cite{asr1}-\cite{asr2} then tells us
 that the configuration average of any function of the random Hamiltonian is a particular matrix element of the same function of $N_i$ as $H$ is of $n_i$.
 
\begin{equation}   \ll  f[H(\{n_i\})] \gg = \langle \{\emptyset\}| \tilde{f}[\tilde{H}(\{\tilde{N}_i\})]|\{\emptyset\}\rangle
\end{equation}
 
 where $|\{\emptyset\}\rangle = \prod^\otimes \left[ \sum_m \sqrt{Pr(\lambda^i_m)}|\lambda^i_m\rangle\right]$  In our alloy model we introduce a random variable $n_i$, one for each muffin-tin sphere,
 such that it takes two values 0 and 1 with probabilities $x$ and $y$ respectively depending on whether a Ni or a Pt ion-core occupies that muffin-tin. The Hamiltonian comes out to be :

 \begin{equation}  
H = \sum_{iL} \left\{ \epsilon^A_L n_i +\epsilon^B_L (1-n_i)\right\} {\rm P}_{iL} +
 \sum_{iL}\sum_{jL'} \left[ t^{AA}_{iL,jL'} n_i n_j + t^{AB}_{iL,jL'} \left( n_i(1-n_j) + n_j(1-n_i)\right) + t^{BB}_{iL,jL'} (1-n_i)(1-n_j)\right] {\rm T}_{iL,jL'}
\label{eqA} \end{equation}      
 here P$_{iL}= |iL\rangle\langle iL|$ and T$_{iL,jL'}= |jL'\rangle\langle iL|$
are projection and transfer operators in the Hilbert space ${\cal H}$ spanned by the LMTO basis $\{|iL\rangle\}$.  The configuration space $\Phi_i$ of $n_i$ is of rank two spanned by  $|\emptyset^i>=\sqrt{x}|0^i> + \sqrt{y}|1^i>$ and $|+^i> = \sqrt{y}|0^i> - \sqrt{x}|1^i>$ and 
 
\[ N_i = x {\rm P}^{\emptyset}_i + y {\rm P}^+_i + \sqrt{xy} \left({\rm T}^{\emptyset +}_i
+{\rm T}^{+\emptyset}_i\right)\] 
 
 Combining the above with Eqn.(\ref{eqA}) we get :

\begin{eqnarray}
\widetilde{H}& = &\sum_{iL} {\rm P}_{iL} \otimes \left[ <\epsilon> \tilde{\rm I} + \epsilon^{(1)} \P +
\epsilon^{(2)} \T \right] + \sum_{iL}\sum_{jL'} {\rm T}_{iL,jL'} \otimes \left[ <t> \tilde{\rm I} +\right.\ldots\nonumber\\
& & \left.+ t^{(1)} (\P + \Q) + t^{(2)} (\T + \TT) + t^{(3)} (\P \otimes \TT + \Q \otimes \T) + t^{(4)} \P\otimes\Q + t^{(5)} \T\otimes\TT \right]
\end{eqnarray}

The next step is to take the augmented Hamiltonian $\widetilde{H}$ and use the Recursion method with
the starting state $|iL\otimes\emptyset\rangle$ to generate a continued fraction expansion of the configuration averaged spectral density or optical response :

\be
 \ll G_{11}(z)\gg  = \langle 1\otimes\emptyset| (z\tilde{I} - \widetilde{H} )^{-1}|1\otimes \emptyset\rangle  =
  \frac{1} {\displaystyle z-\tilde{\alpha}_1 - \frac{\tilde{\beta}^2_2}{\displaystyle z - \tilde{\alpha}_2 -\frac{\tilde{\beta}^2_3}{\displaystyle z-\tilde{\alpha_3} - \frac{\tilde{\beta}^2_3}{\displaystyle  \frac{\ddots}{z -\tilde{\alpha}_N-\tilde{\beta}\tilde{\rm T}(z) }}}}}
 \ee

A large number of publications are available on the determination of the ``terminator" $T(z)$ from
$\{\alpha_1,\ldots \alpha_N\}$ and $\{\beta^2_1,\ldots \beta^2_{N+1}\}$ \cite{term} - \cite{viswa}.  

 For a disordered system, the configuration averaged optical current-current response function is given by

\begin{eqnarray}
C_{\mu\nu}(\vec{r}-\vec{r}^{\ \prime},t)&\ =\  &\ll J_\mu(\vec{r},t)J_\nu(\vec{r}^{\ \prime},0)\gg\\
 \end{eqnarray}

The recursion algorithm is now a change of basis which turns the representation of 
 the current operator into a Jacobian matrix.
 
\begin{eqnarray*}
	   |f_{1}> &\qquad = & \qquad   J_\mu  |1L> \hspace{6cm} \mbox{\rm (initial condition)}\\
 |f_{2}>&\quad\quad =& \qquad\quad H|f_{1}>\quad -\quad \tilde{\alpha}_{1} |f_{1}>\hskip 0.5cm\mbox{\quad\quad and\quad\qquad  }  <f_{2}|f_{1}> =     0 \mbox{\quad\quad if\quad\quad  }\hskip 1cm \tilde{\alpha}_{1} = \frac{<f_{1}|H|f_{1}>}{<f_{0}|f_{0}>} \\         
 |f_{3}>&\qquad =&\qquad H|f_{2}> - \tilde{\alpha}_{2} |f_{2}> -\tilde{\beta}_{2}|f_{1}>\\  <f_{3}|f_{2}>=0  &\mbox{\qquad  gives\qquad }& \tilde{\alpha}_{2} = \frac{<f_{2}|H|f_{2}>}{<f_{1}|f_{1}>} \mbox{\qquad and\qquad  } \tilde{\beta}^2_{1} = \frac{<f_{2}|f_{2}>}{<f_{1}|f_{1}> }\end{eqnarray*}
 The recursive equations described above lead to a  continued fraction 
 expansion of the correlation function :

 \begin{equation}
  {\rm S}(z) = \frac{1}{\displaystyle z-\tilde{\alpha}_1 - \frac{\tilde{\beta}^2_2}{\displaystyle z - \tilde{\alpha}_2 -\frac{\tilde{\beta}^2_3}{\displaystyle z-\tilde{\alpha}_3 - \frac{\tilde{\beta}^2_3}{\displaystyle  \frac{\ddots}{z -\tilde{\alpha}_N-\tilde{\beta}^2_{N+1} {\cal T}(z) }}}}}\\
    \end{equation}
here,  $z=\omega-i\delta$. 

 The correlation function is its Laplace transform. The dielectric function  and optical conductivity are related :
 
\begin{equation}{\rm S}(z) =  \int_0^\infty \ d\tau\ \exp(-z\tau) \ {\rm C}(\tau) \qquad 
  \epsilon_2(\omega) = 4\pi \lim_{\delta\rightarrow 0}\frac{ {\rm S}(z)}{z^2} = \frac{\omega}{4\pi}\sigma(\omega)
\end{equation}

\end{widetext}

\section*{Acknowledgements}
 The authors thank R.Haydock and C.M.M.Nex for permission to use and modify the Cambridge Recursion Package. We would like to thank Prof. O.K. Anderson, Max Plank Institute, Stuttgart, Germany, for his kind permission to use TB-LMTO code developed by his group.

\end{document}